\documentclass{IEEEtran}
\usepackage{amsmath,amsfonts}
\usepackage{algorithm}
\usepackage{algpseudocode}
\usepackage{array}
\usepackage{textcomp}
\usepackage{stfloats}
\usepackage{url}
\usepackage{verbatim}
\usepackage{graphicx}
\usepackage{cite}
\usepackage{tikz}
\usepackage{amssymb}
\usepackage{pgfplots}
\usepackage{textcomp}
\pgfplotsset{compat=newest}
\usepackage{tikz}
\usepackage{amssymb, flushend}
\usepackage{pgfplots}
\usepackage{ifthen}
\usepackage{xcolor} % 引入 xcolor 宏包
\newtheorem{proposition}{Proposition}

\begin{document}

\title{Simultaneous Information and Control Signalling Protocol for RIS-Empowered Wireless Systems}

\author{Evangelos Koutsonas, Xiaonan Mu, Nan Qi, Stylianos Trevlakis, Theodoros A. Tsiftsis, and Alexandros-Apostolos A. Boulogeorgos 
\thanks{E. Koutsonas \& A.-A. A. Boulogeorgos is with the Department Electrical and Computer Engineering, University of Western Macedonia, 50100 Kozani, Greece (e-mail: aboulogeorgos@uowm.gr).}
\thanks{Xiaonan Mu and N. Qi are with  Nanjing University of Aeronautics and Astronautics, Nanjing 210016, China, and also with the State Key Laboratory of Integrated Services Networks, Xidian University, Xi’an 710071, China (e-mail:  nanqi.commun@gmail.com).}
\thanks{S. Trevlakis is with the Department of Research and Development, InnoCube P.C., 55535 Thessaloniki, Greece (e-mail: trevlakis@innocube.org).}
\thanks{T. A. Tsiftsis is with the Department of Informatics and Telecommunications, University of Thessaly, 35100 Lamia, Greece (e-mail: tsiftsis@uth.gr). }
%\affil{Department of Electrical and Computer Engineering, University of Western Macedonia, Kozani, Greece}
%\affil{Research \& Development Department of InnoCube IKE, Thessaloniki, Greece}
%\affil{Department of Informatics and Telecommunications, University of Thessaly, Greece}
%\corresp{CORRESPONDING AUTHOR: A.-A. A. Boulogeorgos (e-mail: aboulogeorgos@ uowm.gr).}
\thanks{The work of E. Koutsonas and A.-A. A. Boulogeorgos was supported by the research project MINOAS. The research project MINOAS is within the H.F.R.I. call ``Basic Research Financing (Horizontal support of all Sciences)" under the National Recovery and Resilience Plan ``Greece 2.0" funded by the European Union - NextGenerationEU (H.F.R.I. Project Number: 15857).}}
%\markboth{IEEE Internet of Things Journal}{A.-A. A. Boulogeorgos, \textit{et al.} Simultaneous information and control signalling protocol for RIS empowered wireless systems}
        % <-this % stops a space
%\thanks{F. Papadimitriou, E. Koutsonas, and A.-A. A. Boulogeorgos are with the }% <-this % stops a space

%\thanks{The work of E. Koutsonas and A.-A. A. Boulogeorgos was supported by the research project MINOAS. }

%\thanks{Manuscript received xxxx, 2023; revised xxx, 2023; accepted xxx, 2023.}

%\markboth{IEEE Wireless Communication Letters, \LaTeX\~Vol.~x, No.~x, xx~2023}%
%{F. Papadimitriou \MakeLowercase{\textit{et al.}}: A non-orthogonal signaling and communication protocol for reconfigurable intelligent surface empowered wireless systems}

%\IEEEpubid{0000--0000/00\$00.00~\copyright~2023 IEEE}

\maketitle

\begin{abstract}
%Reconfigurable intelligent surfaces (RISs) have been recognized as key enablers for reliable and energy efficient wireless systems.
Integration of RIS in radio access networks requires signaling between edge units and the RIS microcontroller (MC). Unfortunately, in several practical scenarios, the signaling latency is higher than the  communication channel coherence time, which causes outdated signaling at the RIS. To counterbalance this, we introduce a simultaneous information and control signaling (SICS) protocol that enables operation adaptation through wireless control signal transmission. SICS assumes that the MC is equipped with a single antenna that operate at the same frequency as the RIS. RIS operates in simultaneous transmission and reflection (STAR) mode, and the source employs non-orthogonal multiple access (NOMA) to superposition the information signal to the control signal. To maximize the achievable user data rate, while ensuring the MC's ability to decode the control signal, we formulate and solve the corresponding optimization problem that returns RIS's reflection and transmission coefficients as well as the superposition coefficients of the NOMA scheme. Our results reveal the robustness of the SICS approach.          
\end{abstract}


\section{Introduction}
\IEEEPARstart{R}{econfigurable} intelligent surfaces (RISs) have emerged in the wireless world as key enablers of high-frequency wireless systems due to their ability to create favorable electromagnetic environments~\cite{9847080}.
%Boulogeorgos2020,Tsiftsis2022,
%Trevlakis2024}. 
An RIS consists of several meta-atoms (MAs) connected through a switching circuit to a controlling unit, which is usually a microcontroller (MC) or a low-complexity field-programmable gate array. The MA can operate in transmit or reflection mode opening the door to the concept of simultaneous transmitting and reflecting (STAR) RIS~\cite{9437234}. The controlling unit is responsible for changing the electromagnetic characteristics of the MAs to cooperatively create the intended~wavefronts~\cite{Tsinos2024}.

In the technical literature, several contributions that quantify and optimize the performance of RIS-empowered systems can be identified~\cite{9352958,10002889,9629335,9815097}
All the aforementioned contributions assume that the MC is connected to the network through an independent low-data-rate link that allows operation control through an application programming interface (API)~\cite{Tsampazi2025}. As a result, operation change, e.g., from beam steering in a specific three-dimensional (3D) angle to another 3D angle or even to beam splitting, requires an important amount of time that depends on the core or edge network latency and may even reach the orders of minutes. This causes important quality of experience degradation and is one of the key barriers of deploying RIS in realistic scenarios. As a response to the need of real-time signaling, a direct communication between the BS and the RIS MC is required. In this direction, the most common approach that was presented contains the establishment of a wired channel, usually through power line communications, between the BS and the RIS controller~\cite{8796365}. However, such an approach limits the flexibility of the installation of RISs and it is not applicable to several realistic scenarios. Motivated by this, the authors of~\cite{SORIS} articulated a novel RIS architecture in which the MC is equipped with a receiver operating at the same frequency as the RIS. The RIS MAs can change their operation mode from reflection to transmission. 

Building upon the architecture, which was presented in~\cite{SORIS}, we introduce a simultaneous information and control signaling (SICS) protocol that allows the RIS to receive operation commands directly from the BS, without interrupting the BS-RIS-UE connection. To make the most out of the system, we formulate a rate maximization problem that returns the optimal reflection and transmission coefficient at each unit cell of the RIS, as well as the power allocation of the NOMA protocol. 
As the optimization problem is non-convex, we treat the independent variables separately and  form two sub-optimal problems. 
%The first one assumes that the reflection and transmission coefficients of the RIS are fixed and returns closed-form expressions for the power allocation coefficients of the non-orthogonal transmission. The second sub-optimal problem considers fixed power allocation coefficients and employs  the interior-point method to calculate the reflection and transmission coefficients. 
By iteratively solving the two problems, we obtain the optimal set of power allocation coefficients, reflection and transmission coefficients. The computational complexity of the proposed optimization strategy is also analyzed. Finally, Monte Carlo simulations are performed to quantify the feasibility of the SICS protocol as well as the performance of the proposed optimization policy.

% The structure of the rest of the paper is the following: Section~\ref{S:SM} reports the system model and the communication protocol. Section~\ref{S:DataRateMaximizationPolicy} articulates the data rate maximization policy. Numerical results accompanied by discussions and engineering insights are presented in Section~\ref{S:NumericalResuls}. Finally, Section~\ref{S:Conclusions} summarizes the paper highlighting the key findings and provided future directions.  

% \textit{Notations}: In what follows, bold letter are used to denote vector or matrices. The operator $\sum_{i=1}^{N}x_i$ returns the sum of $x_i$ with $i=1,2,...,N$. Moreover, $\sqrt{x}$ and $\exp(x)$ stand for the square root and exponential of $x$, respectively. The expected value operator is represented by $\mathbb{E}\left[\cdot\right]$, while the operator $|x|$ returns  the absolute value with $x$ been a scalar. The operator $|\mathbf{x}|$ stands for the geometrical distance of $\mathbf{x}$. The operators $\max\left(z_1, z_2\right)$ and $\rm{rand(1,N)}$ return the maximum value between $z_1$ and $z_2$, and a vector with random values that belongs in $[0, 1]$ of size $1\times N$. The operator $\mathbf{a}^{T}$ returns the transpose of matrix/vector $\mathbf{a}$. Finally, the Hadamard product of matrices $\mathbf{A}$ and $\mathbf{B}$ is denoted as $\mathbf{A}\odot\mathbf{B}$. 

\section{System model \& communication protocol}\label{S:SM}

\begin{figure}
	\centering
	%\hspace{-3cm}
    \scalebox{0.3}{\hspace{-4cm}\input{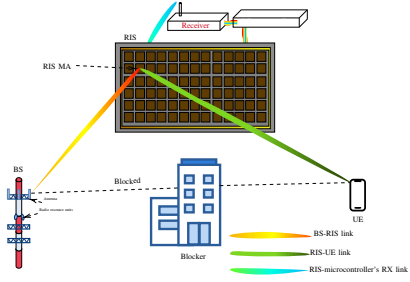}}
	\caption{The considered system model.}
	\label{Fig:SM}
   %\hrulefill 
\end{figure}

 As illustrated in Fig.~\ref{Fig:SM}, we consider a STAR-RIS empowered wireless system, in which the BS communicates through an RIS with a user equipment (UE). The MC is assumed to be equipped with a wireless receiver that operates in the same band as the BS. The RIS employs the STAR scheme in order to allow the MC to capture the transmitted signal. To ensure high energy efficiency, we assume that the BS is equipped with an analog beamformer, the RIS contains $N$ unit-cells. Finally, no direct link is assumed between the BS and UE. 

% \subsection{Communication protocol}
 The communication cycle is assumed to be shorter than the coherence time. The information transmission phase is divided into two phases, namely: (i) simultaneous information and control signaling (SICS), and (ii) data transmission. In the SICS phase, the BS superpositions the information signal to the control signal using NOMA, while the RIS operates in STAR mode. In the data transmission phase, the RIS changes the operation mode and configuration according to the control command that has been previously transmitted in the SICS phase, while the BS continuous the information transmission.     

In the SICS phase, let  $s_{1}, s_{2}$ be the transmitted signals by the BS intended for the UE and the MC, respectively. We assume that NOMA is employed in order for both the control and information signals to reach the corresponding destination. Thus, the baseband equivalent signal at the UE can be obtained~as  
\begin{equation}\label{1a}
 r_{1} =\sum _{i=1}^{N} h_{1i} h_{2i} R_{i} (\sqrt{\alpha _{1} P_{s}} s_{1} +\sqrt{\alpha _{2} P_{s}} s_{2})+n_{1},
\end{equation}
where  
\(\alpha_1,\alpha_2\) denote the power allocation coefficients for the UE and control signal, respectively. The following conditions need to be satisfied for  \(\alpha_1,\alpha_2\): 
%\begin{align}
    $\alpha_1>>\alpha_2$
%\end{align}
and 
%\begin{align}
$\alpha_1+\alpha_2=1.$
%\end{align}
Moreover, 
$R_i = |R_i| \exp\left({j\theta_{R_i}}\right)$ 
where  \(|R_i| \in [0,1]\) denotes the  reflection coefficient magnitudes of the $i$-th RIS unit-cell and \(\theta_{Ri} \in [0,2\pi]\) stand for the phase shifts of the $i$-th RIS unit-cell. $h_{1i}$ and $h_{2i}$ denote the channel coefficients from the BS to $i$-th RIS unit-cell and from $i$-th RIS unit-cell to the UE, respectively. \(P_s\) denotes the BS transmission power. Finally, $n_1$ follows $\mathcal{CN}(0,N_0)$. 

Similarly, the baseband equivalent received signal at the MC can be obtained as 
\begin{equation}\label{eq5}
r_{2} =\sum _{i=1}^{N} h_{1i} g_{i} T_{i} (\sqrt{\alpha _{1} P_{s}} s_{1} +\sqrt{\alpha _{2} P_{s}} s_{2})+n_{2},
\end{equation}
where 
$T_i = |T_i|\,\exp\left({j\theta_{T_i}}\right)$
with \(|T_i| \in [0,1]\) and \(\theta_{T_i} \in [0,2\pi]\) respectively standing for the amplitude and phase of the transmission coefficient. $g_{i}$ denote the channel coefficients from $i$-th RIS unit-cell to the MC. Additionally, $n_2$ follows  the $\mathcal{CN}(0,N_0)$. 

Given that the control command is for beam steering and that the RIS MC was able to correctly detect the command in the SICS phase, in the data transmission phase, the baseband equivalent received signal at the UE can be expressed~as  
    \begin{align}
    r = \sum_{i=1}^{N} \left|R_i^{o}\right| \,\left|h_{1i}\right|\,\left| h_{2i}\right|\,\sqrt{P_s} s + n,
    \end{align}
where $s$ is the transmission symbol and $n$ represents the additive white Gaussian noise of variance $N_0$. Additionally, 
$R_i^{o}=\left|R_i^{o}\right|\,\exp\left({j\theta_{R_i^{o}}}\right)$
is the optimal reflection coefficient of the $i$-th MA. According to~\cite{9095301}, $\left|R_i^{o}\right|\approx 1$ and $\theta_{R_i^{o}}~=~-\theta_{h_{1i}}-\theta_{h_{2i}}$ with $\theta_{h_{1i}}$ and $\theta_{h_{2i}}$ being the phase of $h_{1i}$ and $h_{2i}$, respectively.

\section{Data rate maximization policy}\label{S:DataRateMaximizationPolicy}

% This section formulates and solves the optimization policy that returns the a set of optimal transmission and reflection coefficients, as well as the power allocation, which maximize the achievable data rate at the user equipment, during the SIC phase, while ensuring reliable reception of the control signals at the RIS MC. In this direction, this section is structured as follows: Section~\ref{SS:Data_rate} documents the achievable data-rate, while the optimization problem formulation is given in Section~\ref{SS:ProblemFormulation}. Finally, the data rate maximization policy is presented in Section~\ref{SS:Solution}.  

\subsection{Achievable data rate}\label{SS:Data_rate}
In STAR-RIS NOMA wireless systems, the user employs successive interference cancellation (SIC) to detect the information from the superimposed signal from BS. 
% The BS superimposes the user information to the control message.  The user performs SIC process to recover the information signal, treating the control signal as noise. On the other hand, to decode the control message, which requires a considerably lower data rate than the information signal, the MC's receiver first decodes the information signal, subtracts it from the received signal and then the MC's receivers decodes the control message~\cite{GHAFOOR2022103413}
To achieve this, the following condition should be satisfied: \(\alpha_1>>\alpha_2\).  As a consequence, the SNR at the user can be expressed as 
\begin{equation}
    \gamma_{1} ={\displaystyle \frac{S_1 \alpha _{1} P_{s}}{S_1 \alpha _{2} P_{s} +N_{0} } }. 
    \label{Eq:gamma_1}
\end{equation}
where $S_{1}  \triangleq \left|\sum _{i=1}^{N} h_{1i} h_{2i} R_{i} \right|^{2}$

Assuming a successful SIC at the MC, the SNR at the MC for decoding the information signal and the control message can be respectively obtained~as  
\begin{equation}\label{eq8}
\gamma _{21} ={\displaystyle \frac{S_2 \alpha _{1} P_{s}}{S_2} \alpha _{2} P_{s} +N_{0}}, \,\,
\gamma _{22} ={\displaystyle \frac{S_2 \alpha _{2} P_{s}}{N_{0}}}.
\end{equation}
where $S_{2}\triangleq \left|\sum _{i=1}^{N} h_{1i} g_{i} T_{i} \right|^{2}$.
The achievable data rate at the user is denoted as
$\rho_{1}~=~\log_{2}(1+\gamma_{1})$, while the achievable data rate at the MC can be written~as $\rho_{2} =\log_{2}(1+\gamma_{22})$. 

\subsection{Problem formulation}\label{SS:ProblemFormulation}

The user data maximization in the SICS phase problem can be formulated as
\begin{align}\label{eq11}
    \textbf{P}_1:&\, {\max_{\{\textbf{R}, \textbf{T}\} ,\{\alpha _{1},\alpha_{2}\}} \rho_{1}},  \\ 
    \textbf{s.t.}\,\,
    \label{eq11a} \rm{C}_{1}:&\, |R_{i}|^{2}+|T_{i}|^{2}=1, i=1,2,...,N, \\
    \label{eq11a1}\rm{C}_{2}:&\, |R_{i}|<1, i=1,2,...,N,\\
    \label{eq11a2}\rm{C}_{3}:&\, |T_{i}|<1, i=1,2,...,N,\\
    \label{eq11b} \rm{C}_{4}:&\, \gamma_{21}>\gamma_{th1},\\
    \label{eq11c} \rm{C}_{5}:&\, \gamma_{22}>\gamma_{th2},\\
    \label{eq11d} \rm{C}_{6}:&\, \alpha_{1}+\alpha_{2}=1,\\
    \label{eq11e} \rm{C}_{7}:&\, 0<\alpha_{1}<1,\,\textrm{and}\, 0<\alpha_{2}<1
\end{align}
where, $\textbf{T}=[T_1, T_2,...,T_N],\,
\textbf{R}=[R_1, R_2,...,R_N],$
\(\gamma_{th1}\) and \(\gamma_{th2}\) are the SNR thresholds for decoding the information and control messages at the MC's receiver, respectively. $\rm{C}_{1}$ imposes the energy conservation principle. $\rm{C}_2$ and $\rm{C}_3$ guarantee that both reflection and transmission occurs.  $\rm{C}_4$ and $\rm{C}_5$ ensure that the MC is able to perform SIC and control signal detection, respectively. $\rm{C}_6,\rm{C}_7 $ and $\rm{C}_8$ ensure the power allocation strategy in the NOMA system.

\subsection{Proposed optimization policy}\label{SS:Solution}
Given that $\log_2\left(\cdot\right)$ is an increasing function,  $\textbf{P}_{1}$ can be  written~as
 \begin{align}\label{eq12}
 \textbf{P}_{2}:& \,{\max_{\{\textbf{R}, \textbf{T}\} ,\{\alpha _{1},\alpha_{2}\}}  \gamma_{1}} 
\\
\label{12a}
\textbf{s.t.}  & \,\, \rm{C}_{1}-\rm{C}_{8}. 
\end{align}

Next, we perform independent variables separately to reduce the complexity of the solution policy. Since reflection and ransmission coefficients remain independent of the control signals, we address each problem separately. Therefore, the alternating optimization method is applied to iteratively optimize ($\alpha_1, \alpha_2$) and ($\textbf{R},\textbf{T}$) through the following two propositions:

\begin{proposition}\label{Prop1}
For given $\textbf{R}$ and $\textbf{T}$, the optimal power allocation between the information and control signal can be obtained as
\begin{align}
    \alpha _{1} \ =1-\frac{\gamma_{t h 2} N_{0}}{\left|\sum _{i=1}^{N} h_{1i} g_{i} T_{i} \right|^{2} P_{s}} \,\, \textrm{and}\,\, \alpha _{2} = \frac{\gamma_{t h 2} N_{0}}{\left|\sum _{i=1}^{N} h_{1i} g_{i} T_{i} \right|^{2} P_{s}}. 
    \label{Eq:a1_optimal}
\end{align}
% and
% \begin{align}
%  \alpha _{2} &= \frac{\gamma_{t h 2} N_{0}}{S_{2} P_{s}}. 
%  \label{Eq:a2_optimal}
% \end{align}
\end{proposition}
\begin{IEEEproof}
    The proof of Proposition 1 is given in Appendix~A.   
\end{IEEEproof}

%The following proposition returns the optimization policy for $\left|T_i\right|$ and $\left|R_i\right|$ for fixed $a_1$ and $a_2$. 
%\begin{proposition}
%For given $a_1$ and $a_2$, the optimal values for $\left|T_i\right|$ and $\left|R_i\right|$ are the solutions of the following formulas:
%\begin{equation}
%\sum _{i=1}^{N} |h_{2i} g_{i}| |T_{i}|=\sqrt{\frac{\gamma_{th}^2 N_0}{\alpha_2^2 P_s}}
%\label{Eq:Prop2_1}
% \end{equation}
% and
% \begin{align}
%     \left|R_1\right| = \sqrt{1 -\left|T_i\right|^2}.
% \end{align}
%\end{proposition}
%\begin{IEEEproof}
%    The proof of Proposition 2 is given in Appendix~B.   
%\end{IEEEproof}
%Since \eqref{Eq:Prop2_1} is a linear problem consisting of one equation and $N$ unknown parameters it has infinite solutions. We only need one solution to specify the values of the magnitudes of $T_i$. Moreover, notice that proposition 2 returns the magnitudes od $R_i$ and $T_i$ and not the required phase shifts. The required phase shifts for transmission can be obtained as $\theta_{T_i}=-\theta_{h_{1i}}-\theta_{g_i}$, where $\theta_{g_i}$is are the phase of $g_i$. Similarly, the optimal phase shifts for reflection can be expressed as  $\theta_{R_i}=-\theta_{h_{1i}}-\theta_{h_{2i}}$. 

%\subsection{Proposed optimization policy}
%From Propositions 1 and 2, it becomes evident that in order to determine the optimal values of $\alpha_1$ and $\alpha_2$, it is necessary to know the values of $T_i$. Conversely, to determine the values of $T_i$, knowledge of $\alpha_1$ and $\alpha_2$ is required.
\begin{proposition}\label{Prop2}
For given $\alpha_1$ and $\alpha_2$, the optimal $\textbf{R}$ and $\textbf{T}$ can be obtained by solving the following convex optimization problem:
\begin{equation}
	\begin{aligned}
	 \max _{\textbf{t}}& \sum_{i=1}^N d_i\sqrt{1-{t_i}^2} \\
		 \textbf { s.t. } %&\\
		 & \, 0<t_i<1, \, %i=1,2,...,N, 
         \textrm{and}\,\, \sum_{i=1}^{N}H_it_i=Q,
	\end{aligned}
\end{equation}
where $\textbf{t}=[t_1,t_2,...,t_N],$ $d_i=|h_{1i}||h_{2i}|,$ 
$t_i=|T_i|, $
$|R_i|=\sqrt{1-{t_i}^2},$  
$H_i=|h_{1i}\,g_i|,$ and $
    Q=\sqrt{\frac{{\gamma _{th2}{N_0}}}{{\alpha _2{P_s}}}}.$
This problem is solved using the interior-point method in CVX or other solvers. The required phase shifts for transmission can be obtained as 
$\theta_{T_i}=-\theta_{h_{1i}}-\theta_{g_i},$ where $\theta_{g_i}$ is the phase of $g_i$. Similarly, the optimal phase shifts for reflection can be expressed as $
\theta_{R_i}=-\theta_{h_{1i}}-\theta_{h_{2i}}.$
Therefore, $R_i=|R_i|e^{\theta_{R_i}}$ and $T_i=|T_i|e^{\theta_{T_i}}$, with $i=1,2,...,N$.
\end{proposition}
\begin{IEEEproof}
The proof of Proposition 2 is given in Appendix~B.  
\end{IEEEproof}

By combining Propositions~\ref{Prop1} and~\ref{Prop2}, we present the overall optimization policy, which is described as Algorithm~\ref{alg:Optimal}. Algorithm~\ref{alg:Optimal} takes as inputs the number of iterations, $K$, the initial value of $a_1$, $P_s$, $N_0$, the SNR thresholds, $\gamma_{th1}$ and $\gamma_{th2}$, as well as the number of MAs of the RIS, $N$, and returns the power allocation factors $a_1$ and $a_2$, as well as the reflection and transmission coefficient vectors $\mathbf{R}$ and $\mathbf{T}$. During the initialization phase, there are $\{\alpha_1^{(0)},\alpha_2^{(0)},\textbf{R}^{(0)},\textbf{T}^{(0)}\}\Rightarrow \{\alpha_1^{(0)},\alpha_2^{(0)},S_1^{(0)},S_2^{(0)}\}$.

After the first iteration, the following equation is satisfied:
\begin{equation}
    \begin{aligned}
        \alpha_2^{(1)} = \frac{\gamma_{t h 2} N_{0}}{S_{2}^{(0)} P_{s}}, \,\, \textrm{and} \,\,
        S_2^{(1)} = \frac{\gamma_{th2}N_{0}}{\alpha_{2}^{(1)}P_{s}}
    \end{aligned}.
\end{equation}

% Following the first iteration completion, the derivation is obtained from Propositions~\ref{Prop1}, we derive
%     \begin{align}
%         \frac{\alpha_{1}}{\alpha_{2}}=\frac{1-\alpha_{2}}{\alpha_{2}}, 
%     \end{align}
% which can be bounded as
% \begin{align}
%     \frac{\alpha_{1}}{\alpha_{2}} \geq \frac{\gamma_{th1}}{\gamma_{th2}} + \gamma_{th1},
% \end{align}
% % or equivalently
% % \begin{align}
% %         \gamma_{th2}-\alpha_2\gamma_{th2} &\geq \alpha_2\gamma_{th1}+\alpha_2\gamma_{th1}\gamma_{th2},
% % \end{align}
% which, in turn, leads to 
% % \begin{align}
% %         \frac{\gamma_{th2}}{\alpha_2}&\geq\gamma_{th1}+\gamma_{th2}+\gamma_{th1}\gamma_{th2},
% % \end{align}
% %or
% \begin{align}
%         \frac{\gamma_{th2}}{\alpha_2}\frac{N_0}{P_s} &\geq(\gamma_{th1}+\gamma_{th2}+\gamma_{th1}\gamma_{th2})\frac{N_0}{P_s}\\
%         S_2&\geq(\gamma_{th1}+\gamma_{th2}+\gamma_{th1}\gamma_{th2})\frac{N_0}{P_s}
%     \end{align}

% Therefore, condition~\eqref{con1} must hold, and the next iteration can be carried out.
In the second iteration, the following equation is satisfied:
\begin{equation}
    \begin{aligned}
        \alpha_2^{(2)} 
        = \frac{\gamma_{t h 2} N_{0}}{S_{2}^{(1)} P_{s}}
        %=\frac{\gamma_{t h 2} N_{0}}{\frac{\gamma_{th2}N_{0}}{\alpha_{2}^{(1)}P_{s}} P_{s}}
        =\alpha_2^{(1)}, \textrm{and} \,\,S_2^{(2)} = \frac{\gamma_{th2}N_{0}}{\alpha_{2}^{(2)}P_{s}}=S_2^{(1)}
    \end{aligned}
\end{equation}
% and
% \begin{equation}
%     \begin{aligned}
%         S_2^{(2)} = \frac{\gamma_{th2}N_{0}}{\alpha_{2}^{(2)}P_{s}}=S_2^{(1)}
%     \end{aligned}
% \end{equation}
So after solving problem $\textbf{P}_{3}$ once and problem $\textbf{P}_{7}$ once, the optimization problem can obtain the optimal solution.

\begin{algorithm}
    \caption{Overall optimization policy}\label{alg:Optimal}
    \begin{algorithmic}[1]
    \renewcommand{\algorithmicrequire}{\textbf{Input:}}
    \renewcommand{\algorithmicensure}{\textbf{Output:}}
    \Require\, $K$:\, \textit{Number of iterations}, $a_{1,o}$:\,\textit{Initial value of $a_1$}, $P_s$, $N_0$, $\gamma_{\rm{th1}}$, $\gamma_{\rm{th2}}$, $N$ 
    \Ensure\, $a_1$, $a_2$, $\mathbf{R}$, $\mathbf{T}$ 
    
    \hspace{-36px}\textit{Auxiliary variables:}
    \State $\mathbf{E}=\left[\begin{array}{c c c} 1 & \cdots & 1 \end{array}\right]^{T}$ 

    \hspace{-36px}\textit{Initialization:}
    \State $a_1\gets a_{1,o}$ 
    \State $a_2\gets 1-a_1$  
    \State $\mathbf{R}\gets \rm{rand}(1,N)$ 
    \State $\gamma_1\gets$ Eq.~\eqref{Eq:gamma_1} 
    \State $l_1\gets\frac{\left(\gamma_{\rm{th,1}}\,\gamma_{\rm{th,2}}+\gamma_{\rm{th,1}}+\gamma_{\rm{th,2}}\right)\, N_0}{P_s}$ 
    \State $l_2\gets\gamma_1$ 

    \hspace{-36px}\textit{For loop:}
    \For{$i=1$ to $K$} \do \\
        \If{$\left|\sum _{i=1}^{N} h_{1i} g_{i} T_{i} \right|^{2}\geq l_1$}
            \State $b_1 \gets 1 - \frac{\gamma_{\rm{th2}}\, N_0}{S_2\,P_s}$
            \State $b_2\gets \gamma_{\rm{th1}}\frac{S_2\,P_s+N_0}{S_2\,\left(P_s+\gamma_{\rm{th1}}\right)}$
            \State $a_1\gets b_1$ 
            \State $a_2\gets 1-a_1$
        \EndIf
            \State $\rho_1\gets\log_2\left(1+\gamma_{\rm{th1}}\right)$
        
        \If{$\frac{a_1}{a_2}>b_2$}
            \State $\mathbf{d}\gets 2\left(\left|\mathbf{h_1}\right|^{T}\odot\left|\mathbf{h_2}\right|\right)^{T}$  
            \State $\mathbf{H}\gets 2\left(\left|\mathbf{h_1}\right|^{T}\odot\left|\mathbf{g}\right|\right)^{T}$
            \State $Q_l \gets \sum_{n=1}^N H_n $
            \State $Q_2\gets \sqrt{\frac{\gamma_{\rm{th1}}\,N_0}{P_s\left(a_1-a_2\,\gamma_{\rm{th1}}\right)}}$
            \State $Q\gets 2\max\left(Q=\sqrt{\frac{{\gamma _{th2}{N_0}}}{{\alpha _2{P_s}}}}, Q_2\right)$
            \State $\mathbf{R}\gets$  interior-point method for given $a_1$
        \EndIf
        \State $\mathbf{T}\gets\mathbf{E}-\mathbf{R}$
        \State $\gamma_1\gets$ Eq.~\eqref{Eq:gamma_1} 
    \EndFor \\
    \Return $a_1, a_2,  \mathbf{R}, \mathbf{T}$
    \end{algorithmic}
\end{algorithm}

\subsection{Complexity analysis}
The total number of constraints is $O\left( N \right)$. In interior-point methods, each iteration requires solving a linear system derived from Karush-Kuhn-Tucker conditions. Without exploiting problem structure, the computational complexity for solving this system is $O\left( {{N^3}} \right)$. With approximately $O\left( {\sqrt N \log (1 / \varepsilon)} \right)$ iterations needed, the overall complexity~becomes $
%  O\left( {{N^3}\sqrt N \log\left(\frac{1}{\varepsilon }\right)} \right)~=~
O\left( {{N^{3.5}}\log \left(1 / \varepsilon \right)} \right).  $

\section{Numerical results \& discussion}\label{S:NumericalResuls}

This section verifies the effectiveness of the alternating optimization algorithm and analyzes the impact of key parameters through numerical results. The following scenario is considered where
%: As shown in Fig.~\ref{Simulation Figure 1}, 
the BS, RIS, MC and UE are assumed to be located at (0 m, 0 m), (75 m, 75 m), (75 m, 76 m) and (150 m, 0 m), with heights of 30 m, 10 m, 10 m and 1.5 m, respectively.
% \begin{figure}
% \vspace{-1em}
%     \centering
% %     \includegraphics[width=1\linewidth]{images/Simulation scenario setup.png}  
%     % \caption{Impact of different network structure sizes on algorithm performance}
%     \caption{Simulation scenario setup.}
%     \label{Simulation Figure 1} 
% \end{figure}
The channel model follows the same formulation as in ~\cite{Wangkewei1}, where the RIS-MC link is considered as the RIS-UE link and the BS-UE link is assumed to be completely blocked.
% \begin{figure}
% \vspace{-1em}
%     \centering
% \includegraphics[width=0.5\linewidth]{images/Simulation Figure 1.2.png}  
%     % \caption{Impact of different network structure sizes on algorithm performance}
%     \caption{Convergence of algorithm when $P_{s}=40\text{W}$, $N=64$, $\gamma _{th1} =10\text{dB}$ and $\gamma _{th2}=20\text{dB} $.}
%     \label{Simulation Figure 2} 
% \end{figure}

% Fig.~\ref{Simulation Figure 2} depicts the convergence performance of the alternating optimization algorithm under the following fixed parameters: $P_s = 40\,\text{W}$, $N = 64$, $\gamma_{th1} = 10\,\text{dB}$, $\gamma_{th2} = 20\,\text{dB}$ and $N_0 = -80\,\text{dBm}$. In each iteration, $\alpha_1$ and $\alpha_2$ are optimized through closed-form solutions (\textit{Block 1}), after that $\textbf{R}$ and $\textbf{T}$ undergo iterative optimization using CVX (\textit{Block 2}). The alternating optimization algorithm achieves stable convergence within a single iteration, showing favorable convergence performance. This observation is theoretically validated by Appendix C.
% This figure demonstrates that the algorithm converges to optimal values within few iterations.

\begin{figure}
%\vspace{-1em}
    \centering
\includegraphics[width=0.8\linewidth]{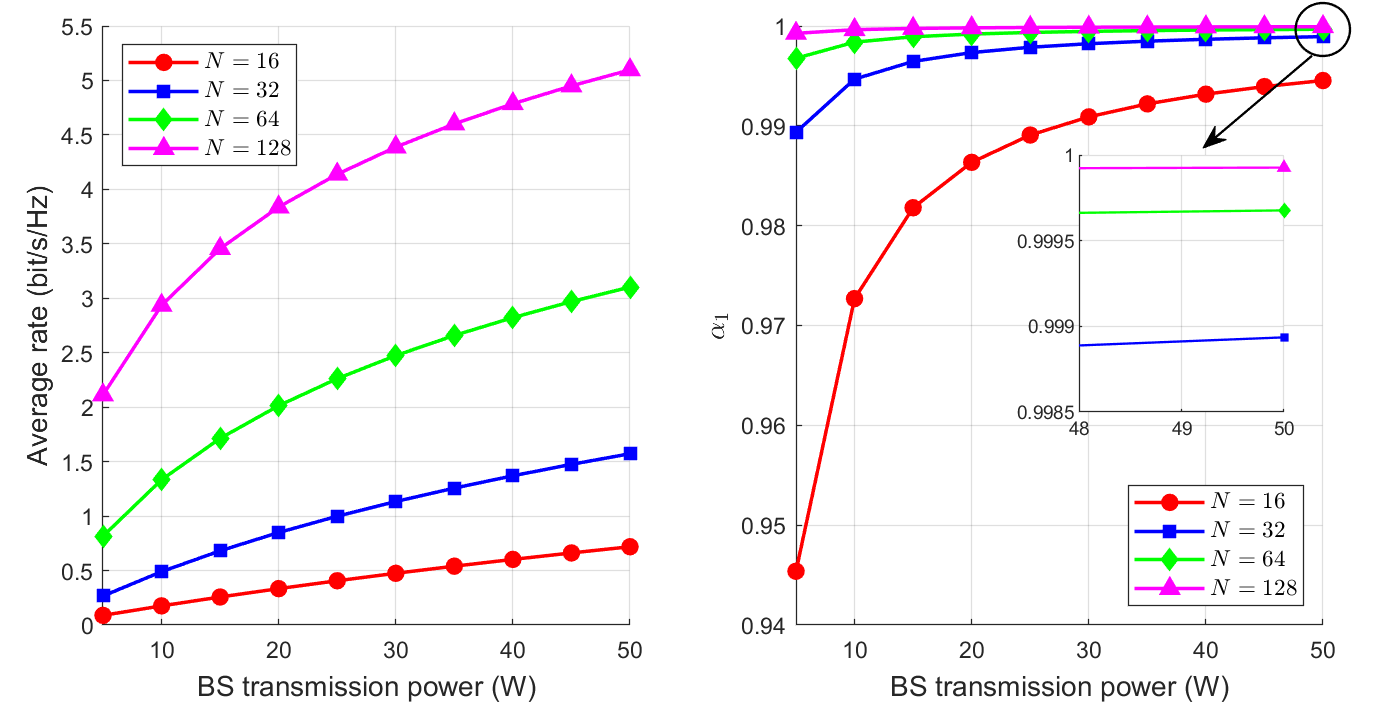}  
    % \caption{Impact of different network structure sizes on algorithm performance}
    \caption{Average rate and $\alpha_1$ vs $P_{s}$ and $N$ when $\gamma _{th1} =10\text{dB}$, $\gamma _{th2}=20\text{dB} $.}
    \label{Simulation Figure 3}  
\end{figure}
Fig.~\ref{Simulation Figure 3} illustrates the UE average rate and $\alpha_1$ as a function of the BS transmit power $P_s$ for different number of RIS MAs $N$. The parameters are set as $\gamma_{th1} =10\,\text{dB}$ and $\gamma_{th2}=20\,\text{dB} $, with each curve averaged over $100$ channel realizations. As expected, for a given $N$, as $P_s$ increases, the average rate increases. Similarly, for a fixed $N$, as $P_s$ increases, $\alpha_1$ also increases. Interestingly, the values of $\alpha_1$ are higher than $0.95$ in all the transmission power regime. This indicates that less than $5\%$ of the transmission power is dedicated to RIS signaling. Moreover, for a given $P_s$, as $N$ increases, the diversity order increases for both the information and signaling channels; thus, the channel conditions improves and consequently the average rate increase. Note that at the high-$P_s$ regime, the performance gain from increasing $N$ diminishes, requiring a trade-off between hardware cost and system performance.

\begin{figure}
%\vspace{-1em}
    \centering
\includegraphics[width=0.8\linewidth]{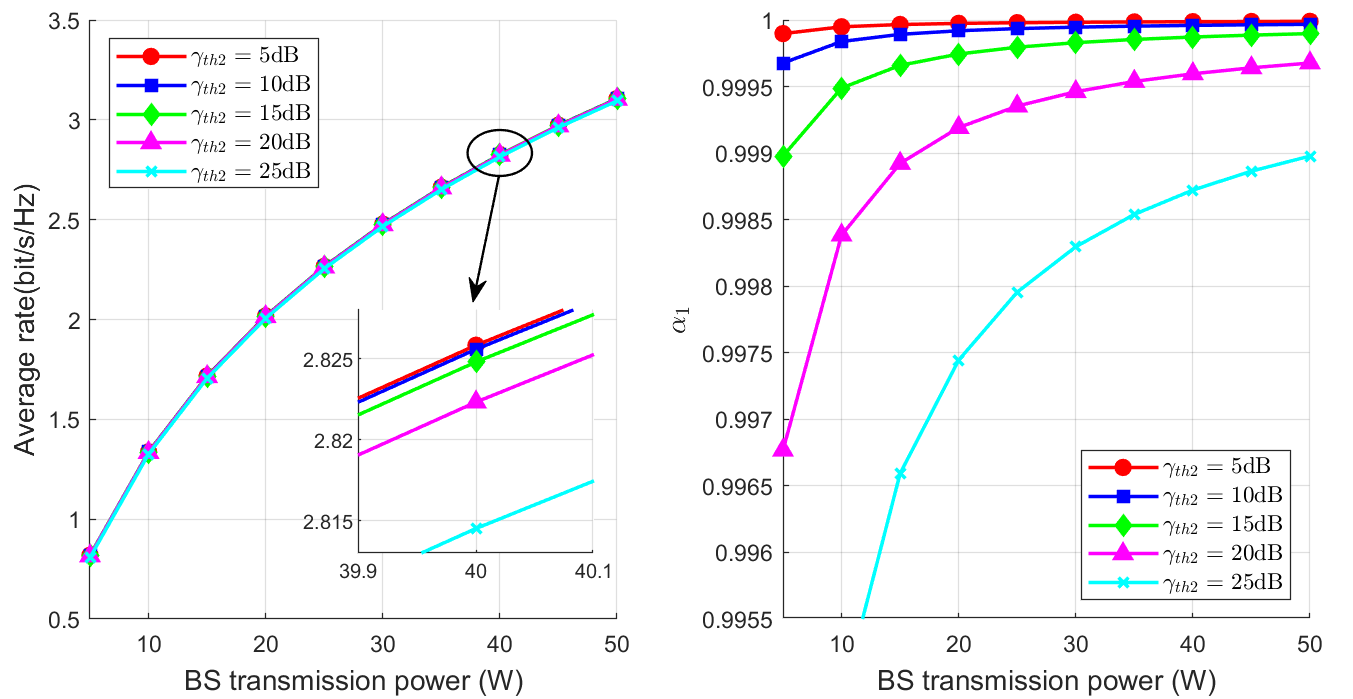}  
    % \caption{Impact of different network structure sizes on algorithm performance}
    \caption{Average rate and $\alpha_{1}$ vs $P_{s}$ and $\gamma _{th2} $ when $\gamma _{th1} =10\text{dB}$, $N=64 $.}
    \label{Simulation Figure 4} 
\end{figure}
Fig.~\ref{Simulation Figure 4} illustrates the UE average rate and $\alpha_1$ as a function of the BS transmit power $P_s$ for different values of $\gamma_{th2} $. The parameters are set as follows: $\gamma_{th1} =10\,\text{dB}$ and $N=64 $, with each curve averaged over $100$ channel realizations. As expected for a given $\gamma_{th2}$, as $P_s$ increases, the average rate also increases. For example, for $\gamma_{th2}=5\,\text{dB}$, as $P_s$ increases from $20$ to $40\,\text{W}$, the average rate increases for approximately $1.8\,\text{bit/s/Hz}$. Additionally, for a fixed $P_s$, as $\gamma_{th2}$ increases, the MC rate increases; thus, the UE average rate decreases. For instance, for $P_s=40\,\rm{W}$, as $\gamma_{th2}$ increases from $15$ to $25\,\rm{dB}$, the average rate decreases from $2.825$ to $2.815\,\rm{bits/s/Hz}$. The significant increase in $\gamma_{th2}$ results in negligible average rate degradation due to the high diversity order achieved by the RIS. From Fig.~\ref{Simulation Figure 4}, we also observe that, for a given $\gamma_{th2}$, as $P_s$ increases, $\alpha_1$ increases, which indicates that $\alpha_2$ decreases. This indicates that for a given control signal rate, decoding the control signal requires a specific amount of received power. Finally, for a fixed $P_s$, as $\gamma_{th2}$ increases, the data rate of the control signal increases; as a consequence, the received power at the MC needs to increases and in turn $\alpha_1$ decreases.

\section{Conclusions} \label{S:Conclusions}

In this work, a SICS protocol was presented that minimizes the signaling latency between the BS and the RIS. In more detail, the protocol was based in two concepts: i) NOMA and ii) STAR operation of the RIS. %NOMA was used to simultaneously transmit information and control messages, while the STAR operation of the RIS was employed to enable simultaneous reception of the transmitted signal by both the MC and the final destination.% 
To make the most out of the presented protocol, we formulate and solve an information data rate maximization problem. The problem is non-convex. To solve it, we treated the independent variables separately and formed two sub-optimal problems. By iteratively solving the two problems, we obtained the optimal set of power, reflection and transmission coefficients. Additionally, we performed complexity analysis and showed that the solution is of low-complexity, i.e., the the proposed policy is feasible. This finding was validated through extensive Monte Carlo simulations. The aforementioned simulations highlighted the robustness and efficiency of the SICS protocol.

\section*{Appendix A}
\section*{Proof of Proposition 1}

Given that \(\alpha_2=1-\alpha_1\), $\textbf{P}_{2}$ can be rewritten~as
\begin{align}
     \textbf{P}_{3}:&\,  {\max_{\alpha_1} \gamma _{1} ={\frac{S_1\alpha_{1}P_{s}}{S_1(1-\alpha_{1})P_{s} +N_{0}}}} \\
    \textbf{s.t.}  \,\,
    \rm{C}_{4}:&\, \frac{S_{2}\alpha_{1}P_{s}}{S_{2}(1-\alpha _{1})P_{s}+N_{0}}>\gamma_{th1},\\
    \rm{C}_{5}:&\, \frac{S_{2}(1-\alpha _{1})P_{s}}{N_{0}}>\gamma_{th2},\\
    % \rm{C}_{6}:&\, \alpha_1+\alpha_2=1. 
    \rm{C}_{7}:&\, 0<\alpha_{1}<1.
\end{align}

Constraints $\rm{C}_1$, $\rm{C}_2$, $\rm{C}_3$ can be neglected as $R_i$ and $T_i$ are fixed. 
Both \(\gamma_1\) and \(\gamma_{21}\) are increasing functions of \(\alpha_1\), while \(\gamma_{22}\) is a decreasing function of \(\alpha_1\). To satisfy $\rm{C}_4$ and $\rm{C}_5$, the following conditions should be valid: 
\begin{equation}
    \alpha_{1}>\frac{\gamma_{t h 1} S_{2} P_{s}+\gamma_{t h 1} N_{0}}{S_{2} P_{s}+\gamma_{t h 1} S_{2} P_{s}} \,\, \textrm{and} \,\,\alpha _{1}<1-\frac{\gamma_{th2}N_{0}}{S_{2}P_{s}}.
\end{equation}
% and
% \begin{equation}
%     \alpha _{1}<1-\frac{\gamma_{th2}N_{0}}{S_{2}P_{s}}\
% \end{equation}
% and 
% \begin{equation}
%     0<\alpha_1<1
% \end{equation}

If and only if $ 1-\frac{\gamma_{t h 2} N_{0}}{S_{2} P_{s}} \geq \frac{\gamma_{t h 1} S_{2} P_{s}+\gamma_{t h 1} N_{0}}{S_{2} P_{s}+\gamma_{t h 1} S_{2} P_{s}}$ is satisfied, that is, when
\begin{equation}\label{con1}
    \begin{aligned}
    S_{2}\geq(\gamma_{th1}\gamma_{th2}+\gamma_{th1}+\gamma_{th2})\frac{N_{0}}{P_{s}}
    \end{aligned}
\end{equation}
can there be
\begin{equation}
    \begin{aligned}
    \alpha_{1} \in (\frac{\gamma_{t h 1} S_{2} P_{s}+\gamma_{t h 1} N_{0}}{S_{2} P_{s}+\gamma_{t h 1} S_{2} P_{s}},1-\frac{\gamma_{t h 2} N_{0}}{S_{2} P_{s}})
    \end{aligned}
\end{equation}

As \(\gamma_{1}\) is an increasing function of $\alpha_1$, \(\alpha_1\) needs to attain the maximum possible value, which leads to~\eqref{Eq:a1_optimal}. 
%and straightforwardly to~\eqref{Eq:a2_optimal}. 
This concludes the proof. 

\section*{Appendix B}
\section*{Proof of Proposition 2}
$\textbf{P}_{2}$ can be rewritten as 
\begin{align}
    \textbf{P}_{4}:\,& {\max_{\{\textbf{R}, \textbf{T}\} } \gamma_{1}={\frac{S_1\alpha_{1}P_{s}}{S_1\alpha_{2}P_{s} +N_{0}}}},  \\ 
    \textbf{s.t.}\,\, %& 
    \rm{C}_{1}:& \, |R_{i}|^{2} +|T_{i}|^{2} =1, i=1,2,...,N,\\
    \rm{C}_{2}:& \, |R_{i}| < 1, i=1,2,...,N,\\
    \rm{C}_{3}:& \, |T_{i}| < 1, i=1,2,...,N,\\
    \rm{C}_{4}:& \, \frac{S_2\alpha_{1}P_{s}}{S_2\alpha_{2}P_{s}+N_{0}}>\gamma_{th1},\\
    \rm{C}_{5}:& \, \frac{S_2\alpha_{2}P_{s}}{N_{0}}>\gamma_{th2},
\end{align}
or equivalently
\begin{align}
    \textbf{P}_{4}:\,& {\max_{\{\textbf{R}, \textbf{T}\} } |\sum_{i=1}^N h_{1i}h_{2i}R_{i}|},  \\ 
    \textbf{s.t.}\,\, %& 
    \rm{C}_{1}:& \, |R_{i}|^{2} +|T_{i}|^{2} =1, i=1,2,...,N,\\
    \rm{C}_{2}:& \, |R_{i}| < 1, i=1,2,...,N,\\
    \rm{C}_{3}:& \, |T_{i}| < 1, i=1,2,...,N,\\
    \rm{C}_{4}:& \, |\sum_{i=1}^N h_{1i}g_{i}T_{i}|>\sqrt{\frac{\gamma_{th1}N_{0}}{\alpha_{1}P_{s}-\gamma_{th1}\alpha_{2}P_{s}}},\\
    \rm{C}_{5}:& \, |\sum_{i=1}^N h_{1i}g_{i}T_{i}|>\sqrt{\frac{\gamma_{th2}N_{0}}{\alpha_{2}P_{s}}}.
\end{align}

Among them, constraint $\rm{C}_{4}$ holds only when $\gamma_{th1}<\alpha_1/\alpha_2$ is satisfied, otherwise problem $\textbf{P}_{4}$ has no solution.

After solving problem $\textbf{P}_{3}$, transform $\alpha_{1}/\alpha_{2}$ based on ~\eqref{con1}, it can be easily proved that $\alpha_{1}/\alpha_{2}>\gamma_{th1}.$
% \begin{equation}
% \begin{aligned}
%     \frac{\alpha_{1}}{\alpha_{2}} 
%     % &=\frac{1-\frac{\gamma_{th2}N_{0}}{S_{2}P_{s}}}{\frac{\gamma_{th2}N_{0}}{S_{2}P_{s}}}\\
%     % &=\frac{S_{2}P_{s}-\gamma_{th2}N_{0}}{\gamma_{th2}N_{0}}\\
%     % &\geq\frac{(\gamma_{th1}\gamma_{th2}+\gamma_{th1}+\gamma_{th2})N_{0}-\gamma_{th2}N_{0}}{\gamma_{th2}N_{0}}\\
%     % &=\frac{\gamma_{th1}}{\gamma_{th2}}+\gamma_{th1}\\
%     >\gamma_{th1}.
% \end{aligned}
% \end{equation}
The latter can definitely be satisfied, and subsequently, 
%\begin{align}
    $\frac{\alpha_{1}}{\alpha_{2}}\geq\frac{\gamma_{th1}}{\gamma_{th2}}+\gamma_{th1},$
%\end{align}
which can be equivalently written as
% \begin{align}
%     \alpha_{1}\gamma_{th2}&\geq\alpha_{2}\gamma_{th1}+\alpha_{2}\gamma_{th1}\gamma_{th2},
% \end{align}
% or
% \begin{align}
%     \alpha_{1}\gamma_{th2}-\alpha_{2}\gamma_{th1}\gamma_{th2}&\geq\alpha_{2}\gamma_{th1},
% \end{align}
% or equivalently 
% \begin{align}
%     (\alpha_{1}-\alpha_{2}\gamma_{th1})\gamma_{th2}&\geq\alpha_{2}\gamma_{th1},
% \end{align}
%that, in turn, leads to
% \begin{align}
%     \frac{\gamma_{th2}}{\alpha_{2}}&\geq\frac{\gamma_{th1}}{\alpha_{1}-\alpha_{2}\gamma_{th1}}
% \end{align}
or
%\begin{align}
    $\frac{\gamma_{th2}}{\alpha_{2}}\frac{N_{0}}{P_{s}}\geq\frac{\gamma_{th1}}{\alpha_{1}-\alpha_{2}\gamma_{th1}}\frac{N_{0}}{P_{s}},$
%\end{align}
which yields
%\begin{align}
    $\sqrt{\frac{\gamma_{th2}N_{0}}{\alpha_{2}P_{s}}}\geq\sqrt{\frac{\gamma_{th1}N_{0}}{\alpha_{1}P_{s}-\gamma_{th1}\alpha_{2}P_{s}}}.$
%\end{align}
Therefore, in $\textbf{P}_{4}$, $\rm{C}_{5}$ is a stronger constraint than $\rm{C}_{4}$, the constraint $\rm{C}_{4}$ can be omitted. By accounting for
%\begin{equation}
%    \begin{aligned}
       $ \left|\sum_{i=1}^N h_{1i}h_{2i}R_{i}\right| = \left|\sum_{i=1}^N |h_{1i}||h_{2i}||R_{i}|\right| = \sum_{i=1}^N |h_{1i}||h_{2i}||R_{i}|$
 %   \end{aligned}
%\end{equation}
and
%\begin{equation}
   % \begin{aligned}
      $ \left|\sum_{i=1}^N h_{1i}g_{i}T_{i}\right| = \left|\sum_{i=1}^N |h_{1i}||g_{i}||T_{i}|\right| = \sum_{i=1}^N |h_{1i}||g_{i}||T_{i}|,$
  %  \end{aligned}
%\end{equation}
$\textbf{P}_{4}$ can be rewritten as 
\begin{align}
    \textbf{P}_{5}:\,& {\max_{\{\textbf{R}, \textbf{T}\} } \sum_{i=1}^N |h_{1i}||h_{2i}||R_{i}|},  \\ 
    \textbf{s.t.}\,\, %& 
    \rm{C}_{1}:& \, |R_{i}|^{2} +|T_{i}|^{2} =1, i=1,2,...,N,\\
    \rm{C}_{2}:& \, |R_{i}| < 1, i=1,2,...,N,\\
    \rm{C}_{3}:& \, |T_{i}| < 1, i=1,2,...,N,\\
    \rm{C}_{5}:& \, \sqrt{S_2}=\sum_{i=1}^N |h_{1i}||g_{i}||T_{i}|>\sqrt{\frac{\gamma_{th2}N_{0}}{\alpha_{2}P_{s}}}.
\end{align}
We noticed the objective function $\sum_{i=1}^N |h_{1i}||h_{2i}||R_{i}|$ is a decreasing function of $|T_{i}|$, therefore, constraint $\rm{C}_5$ can be written in a stronger form as follows
\begin{equation}
    \begin{aligned}
        \sqrt{S_2}=\sum_{i=1}^{N}|h_{1i}||g_{i}||T_{i}|&=\sqrt{\frac{\gamma_{th2}N_{0}}{\alpha_{2}P_{s}}}.
    \end{aligned}
\end{equation} 

$\textbf{P}_{5}$ can be rewritten as 
\begin{equation}
	\begin{aligned}
	\mathbf{P}_{6}: \max _{\textbf{T}}& \sum_{i=1}^N |h_{1i}||h_{2i}|\sqrt{1-|T_i|^2} \\
		 \textbf { s.t. } %&\\
		 & \, 0<|{T}_i|<1, i=1,2,...,N, \\
		 & \, \sum_{i=1}^{N}|h_{1i}||g_{i}||T_{i}|=\sqrt{\frac{\gamma_{th2}N_{0}}{\alpha_{2}P_{s}}}.
	\end{aligned}
\end{equation}

Let $|h_{1i}||h_{2i}|=d_i, \left| {{T_i}} \right| = {t_i}, |h_{1i}||g_i|=|h_{1i}g_i| = {H_i}, \sqrt {\frac{{\gamma _{th2}{N_0}}}{{\alpha_2{P_s}}}}=Q$,
then
\begin{equation}
	\begin{aligned}
	\mathbf{P}_{7}: \max _{\textbf{t}}& \sum_{i=1}^N d_i\sqrt{1-{t_i}^2} \\
		 \textbf { s.t. } %&\\
		 & \, 0<t_i<1, i=1,2,...,N, \\
		 & \, \sum_{i=1}^{N}H_it_i=Q.
	\end{aligned}
\end{equation}

For $t_i,\forall i=1,2,...,N,$ $
    f({t_i}) = \sqrt {1 - {t_i}^2} < 0, \,\,
    {f^{'}}({t_i}) =  - \frac{{{t_i}}}{{\sqrt {1 - {t_i}^2} }}>0, \,\,
    {f^{''}}({t_i}) =  - \frac{1}{{{{\left( {1 - {t_i}^2} \right)}^{\frac{3}{2}}}}} < 0$.
 The superposition of multiple concave functions $t_i$ preserves concavity. Consequently, $\textbf{P}_{7}$ remains convex. This problem is solved using the interior-point method in CVX or other~solvers.

\bibliographystyle{IEEEtran}
\bibliography{refs} 

\end{document}